# General Circulation and Principal Wave Modes in Andaman Sea from Observations

**S.R. Kiran***

Center for Atmospheric and Oceanic Sciences, Indian Institute of Science, CV Raman Road, Devasandra Layout Bangalore – 560012, Karnataka, India; kiransr1991@gmail.com

## Abstract

**Objectives:** This study intends to describe the Andaman Sea circulation and investigate the dominant modes of variability in the basin. **Analysis/Observations:** The domain experiences stronger South-westerly winds from May to September and relatively weaker North-easterlies from November to February. A strong negative Ekman pumping along the north coast of Indonesia is observed during Summer. The transport of water across the straits of Andaman and Nicobar Islands (ANI) is computed by simple mass balance and is found to be in phase with the monthly averaged Mean Sea Level Anomalies (MSLA) of Andaman Sea. There occurs high surface outflux through Great channel and Ten-Degree channel in Summer. In April and October, rate of transport through the straits is maximum. During the same months, meridional surface currents intensify along the eastern boundary of the basin and are associated with signatures of downwelling. **Findings:** Intense down welling occurs to the north coast of Indonesia during Summer, locally forced by south-westerlies. There occurs large influx of water into Andaman Sea between April and November through the straits between the islands. Equatorial Wyrtki jets remotely force Kelvin waves of downwelling nature in the basin during April and October. The circulation in Andaman Sea is characterised by gyres or vortices, which is the manifestation of Rossby waves of semi-annual mode. The basin has a characteristic 120-day mode of westward propagating long Rossby wave packets which reflect from the coast of ANI as eastward propagating short Rossby waves. **Novelty:** The first attempt to completely describe the dynamics of Andaman Sea circulation exclusively from observations.

## 1. Introduction

Andaman Sea (A-SEA), which extends over 92E to 100E and 4N to 20N, occupies a very significant position in Indian Ocean, yet remained unexplored for long period of time. Located to the south of Myanmar, west of Thailand and north of Indonesia, the A-SEA is separated from Bay of Bengal by the Andaman and Nicobar Islands (ANI) and an associated chain of sea mounts along the Indo-Burmese plate boundary. The Strait of Malacca (between Malay Peninsula and Sumatra) forms the southern exit way of the basin, which is 3km wide and 37m deep.

Previous studies[1–2] show that sea level in the basin is high during Summer monsoon due to piling up of water by Ekman drifts. The region experienced heavy rainfall from May to November with an annual average rainfall of 2000mm. The main source of fresh water into A-SEA is from Irrawady, Salween and Thanintharyi rivers, which possess maximum runoff during Summer. The domain experiences two salinity lows in a year; one during September-October months due to freshwater influx

**Author for correspondence*

from rivers and the other during North-East monsoon due to influx from Strait of Malacca. Potemra[3], using ocean models, concluded the flow in the upper ocean layer to be cyclonic in spring and early summer, while it is anti-cyclonic during rest of the year. He observed that Rossby waves generated from A-SEA gets blocked by ANI. Internal waves[4–6] were observed in the basin, which is characteristic of a stratified ocean with disturbances caused by tidal flow passing over shallow water obstacles.

## 2. Data and Methodology

ETOPO5 (5-minute gridded global relief data) is used to study bathymetry of A-SEA. Advanced Scatterometer (ASCAT) data is used to study the surface wind field as well as compute the windstress curl over the region. Monthly averaged climatologies of temperature and salinity with depth are obtained from World Ocean Atlas (WOA) 2013. Sea surface currents are obtained from the Ocean Surface Current Analysis Real-time (OSCAR) estimations with a temporal resolution of 5 days. The Mean Sea Level anomalies (MSLA) altimetry data (with a temporal resolution of 7 days) from Archiving, Validation and Interpretation of Satellite Oceanographic (AVISO) data portal are used to study the ocean circulation and temporal variability of Sea Surface with respect to Mean Sea Level (MSL). Fresh water influx (River Irrawady) into A-SEA is obtained from Global Runoff Data Centre (GRDC)[7]. Rainfall data is obtained from Tropical Rainfall Measuring Mission (TRMM).

In this study, depth corresponding to 20 degree Celsius isotherm is chosen as the proxy for thermocline depth, the shoaling and deepening of which gives insight into upwelling and downwelling events respectively. The width of the straits between ANI is determined by distance computation method in coordinate geometry, where the geographic coordinates of the north and south coasts of the islands are known. Computation of the curl of wind stress involves spatial derivatives of the wind stress, which is performed by means of simple finite-backward differencing method. To study the dominant modes of variability, techniques such as Wavelet Transforms[8] are used to express the geophysical data in both time and frequency domain simultaneously in Figure 9. A black solid line scribing the wavelet spectra is the cone of influence, which gives the maximum period up to which the spectrum is significant.

## 3. Analysis, Results and Discussions

### 3.1 Bathymetry and Straits

The fluid dynamics of an ocean basin is greatly determined by its bathymetry. The first snapshot of Figure 1 shows the sea floor relief of A-SEA measured as the depth with respect to mean sea surface level. The northern and eastern side of the basin is shallow, as the continental shelf off the coast of Myanmar and Thailand extends over 200km (marked by 300m isobath). Figure 2, it is evident that about 45 % of the basin area is shallower (less than 500m depth), which is the direct consequence of the presence of the wider shelf. The continental slope which follows the eastern shelf is quite steep between 9N and 14N (the second snapshot of Figure 1). Here, the perspective view of the submarine topography sectioned along 95E exposes the abrupt rise in depth of sea by about 3000m within a short horizontal distance of a degree. Isobaths corresponding to 900m and 2000m are also shown in the figure to emphasize the steepness of the slope. Further, it may be noted that the deep ocean is also not free from sea mounts; hence only around 15% of the total area is deeper than 2500m (Figure 2).

The western boundary of A-SEA is marked by volcanic islands and sea mounts, with straits or passages of variable depths that control the entry and exit of water to Bay of Bengal. There occurs a drastic change in depth of water over a small distance of 200km, as one moves from Bay of Bengal (around 3500m deep) to the vicinity of islands (upto 1000m depth) and further into A-SEA. The exchange of water between A-SEA and Bay of Bengal occurs through the straits formed between ANI, the width of which as computed are listed in Table 1. Out of these, the most important straits (in terms of width and depth) are: Preparis Channel (PC), Ten Degree Channel (TDC) and Great Channel (GC), across which major transport shall take place and are indicated in Figure 1. PC is the widest but shallowest (250m) of the three and separates South Myanmar from North Andaman. TDC is 600m deep and lies between Little Andaman and Car Nicobar. GC is 1500m deep and separates Great Nicobar from Banda Aceh. These may be considered as the major passages which connect A-SEA to Bay of Bengal and Equatorial Indian Ocean.

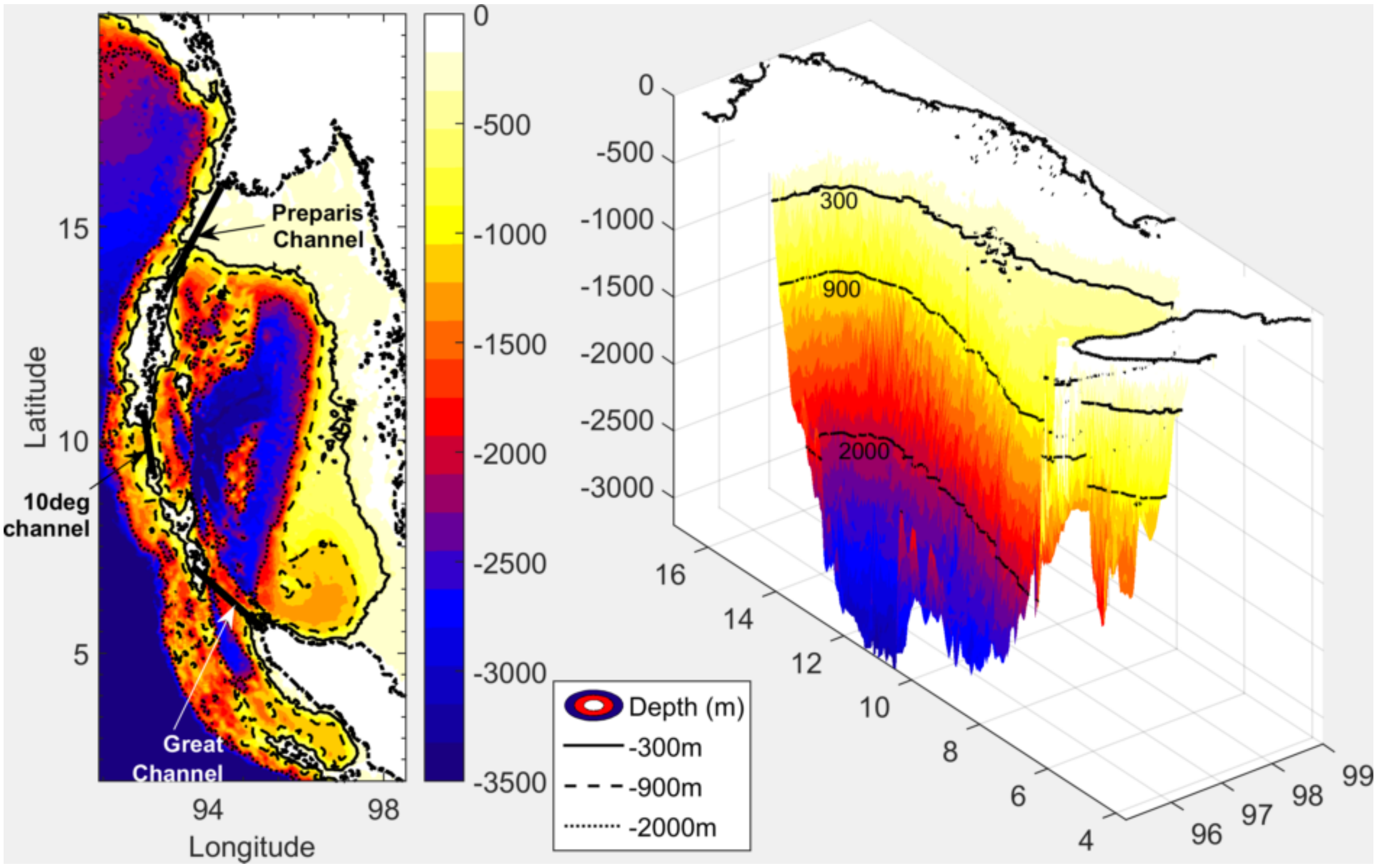

**Figure 1.** The Bathymetry (in metres) of A-SEA in 2D and 3D (sectioned along 95E).

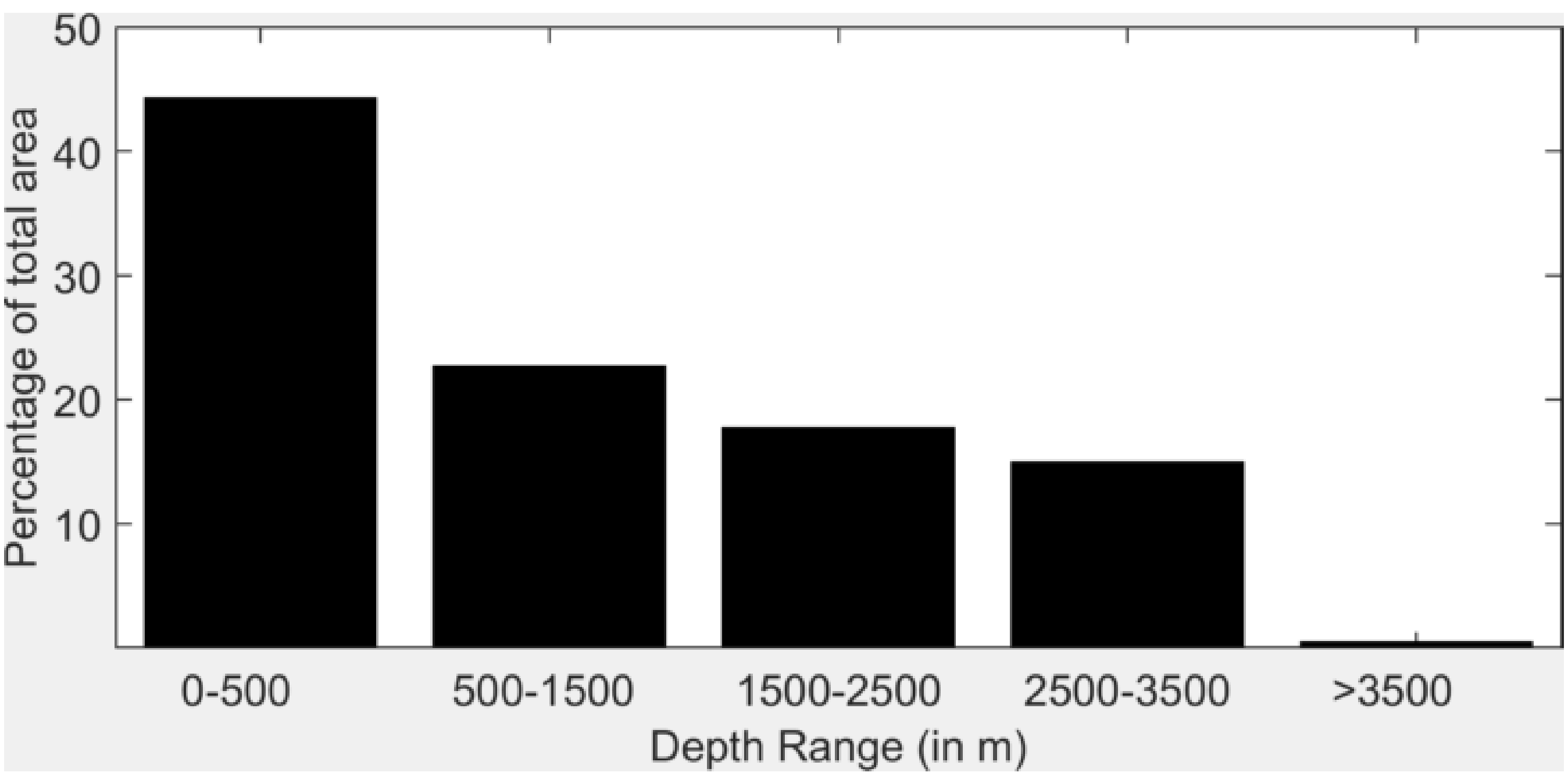

**Figure 2.** Percentage of total area of A-SEA corresponding to different ranges of depth.

### 3.2 Winds

As observed in Indian Ocean, the wind system over A-SEA regime also reverses every year. The monthly averaged ASCAT winds for 2011 (Figure 3) show that the region experiences north-easterlies with an average windspeed of 5 m/s in the months of November to February. During these months, the western part of the domain experiences maximum wind intensity. It weakens by March-April and reverses to strong south-westerlies from May to September, with mean wind speeds touching 8 m/s in June, July and August, distributed nearly uniform over the entire basin. The wind plummets by October and switch back to north-easterlies from November.

The effect of wind stress on ocean surface is explained with the help of wind stress curl. The net divergence of water in the ocean mixed layer results in Ekman Pumping, the vertical velocity of which is given by:

$$W \;\; = \frac{1}{\rho f} \left[ \, curl(\Gamma) + \frac{\beta \Gamma_x}{f} \, \right]$$

where Γ is the horizontal wind stress in the domain with $\Gamma_x$ as its zonal component, ρ is the mean density of sea water, $\beta = (\delta f/\delta y)$ and f is the Coriolis parameter, which is a function of latitude. Here, Γ is obtained from ASCAT wind stress and ρ is taken as 1030 kg/m$^3$, and the monthly averaged pumping velocities are evaluated across the basin. Ekman pumping for the months of June and December are shown in Figure 4. The comparison between the two seasons elicits a very strong negative pumping velocity of more than 5m/day along the north coast of Indonesia from May to September (shown here, June). This signifies a probable tendency of coastal downwelling in Summers. Besides, it is also observed that the region develops a weak but positive pumping velocity (less than 3m/day) at the mouth of GC in Winters (here, December).

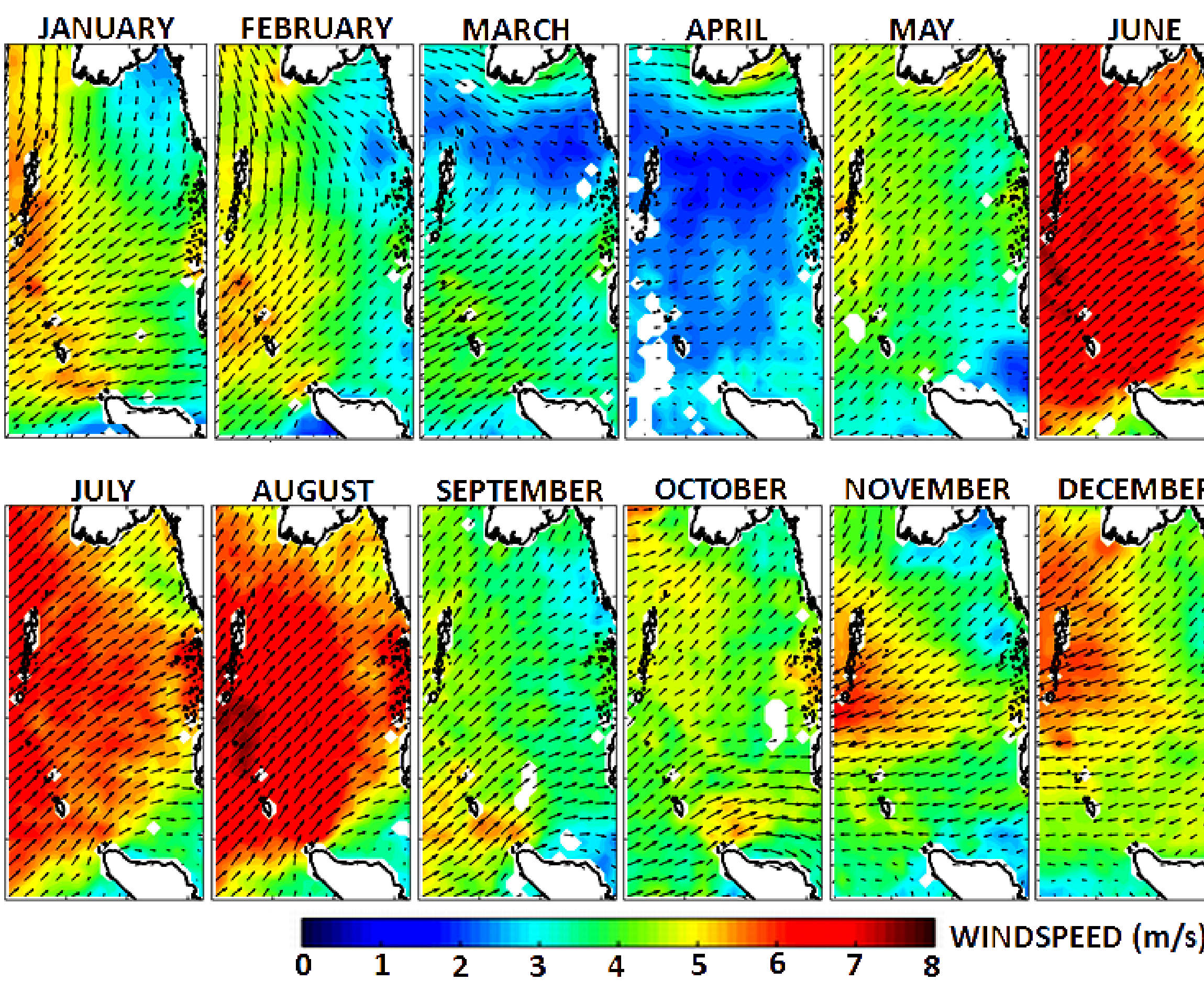

**Figure 3.** Monthly averaged ASCAT winds (in m/s) of A-SEA for the year 2011.

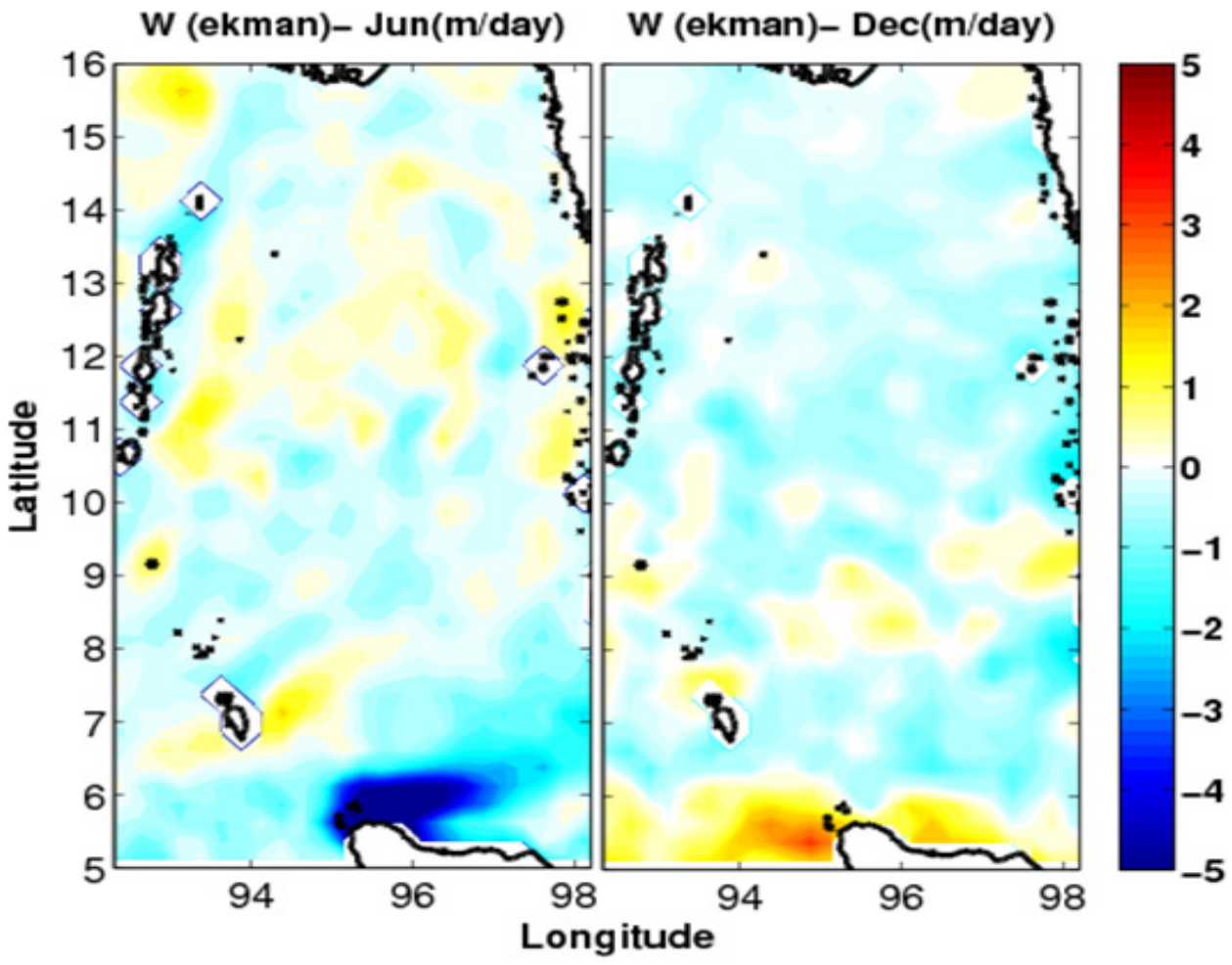

**Figure 4.** Monthly averaged Ekman Pumping velocity (m/day) for June and December.

Depth of 20 degree isotherm averaged over the zonal range of 95E to 96E, determined from WOA Temperatures, is shown in Figure 5. Generally, the depths are found to be maximum between May and December, and minimum between January and April. There is a definite indication of formation of a cell of strong downwelling to the north of Indonesia, which is marked by the deepening of isotherm (depth exceeding 130m). This originates near the coast in May and spreads further north (upto 9N) during June, July and August. A strong negative Ekman pumping (discussed in previous paragraph) during the same period suggests the profound impact of local winds in A-SEA during Summer. Shoaling of isotherms occur between January and March, along the entire band of latitudes, although the effect of wind is pronounced closer to the coast.

## 3.3 Surface Circulation

OSCAR surface currents for 7 years (2007-2013) are averaged to obtain mean monthly circulation of A-SEA (Figure 6). Generally, the currents are found to be stronger at the south than any other part of the basin. An intense surface outfluxes through GC, of the order of 40cm/s occurs during Summers and Winters. While this flow is directed westwards in winter, it is southwards along the west coast of Indonesia in Summer. On the other hand, the TDC has strong surface influx in Summer, which weakens by October. This is followed by a sturdy out flux in winter, which wanes by the month of April. Although the surface flow through PC is generally inward during Summer monsoon, the preceding and succeeding months experience outflow (strong outflow in October, but weak outflow in April).

The intense momentum flux through the straits induces strong shear in the flow and thereby resulting in the formation of vortices in the basin, which are either transient and non-periodic, or persistent and periodic. Only the latter is considered for the present study owing to its geophysical relevance. Two gyres, one cyclonic (at the north) and another anticyclonic (at the south), form in A-SEA during May-June and November-December. The circulation changes its polarity during other months, i.e., an anticyclonic vortex and a cyclonic vortex form during August-September and February-March, at the north and south of the basin respectively (Figures not shown here). The circulation is generally weak in January, April, July and October. Hence, the general circulation is characterised by vortices at the north and south of the domain with alternating polarity of semi-annual period.

During April and October, when the effects of local winds are minimal, A-SEA experiences the intensifica-tion of meridional surface currents in the poleward direction along the continental slope on the eastern side of the basin, with a spatial maxima at 8N, 97E (Figure 6). This is characteristic of the propagation of Kelvin Waves, investigations on which are discussed in the Section 3.5.

**Table 1.** Width of Straits of Andaman and Nicobar Islands

| Sl. No. | Strait | Width (km) | Sl. No. | Strait | Width (km) |
|---|---|---|---|---|---|
| 1 | Preparis Channel | 322 | 9 | Car Nicobar–BattiMalv Island | 30 |
| 2 | Mac. Pherson Strait | 4 | 10 | BattiMalv Island–Chawra Island | 47 |
| 3 | Rutland Island–Cinque Island | 20 | 11 | Chawra Island–Teressa | 13 |
| 4 | Cinque Island–Passage Island | 7 | 12 | Teressa–Katchal | 30 |
| 5 | Passage Island–Sisters Island | 8 | 13 | Nancowry–Katchal | 6 |
| 6 | Sisters Island–Brother Island | 18 | 14 | Katchal–Little Nicobar | 55 |
| 7 | Brother Island-Little Andaman | 16 | 15 | Little Nicobar–Great Nicobar | 18 |
| 8 | Ten degree Channel | 146 | 16 | Great Channel | 168 |

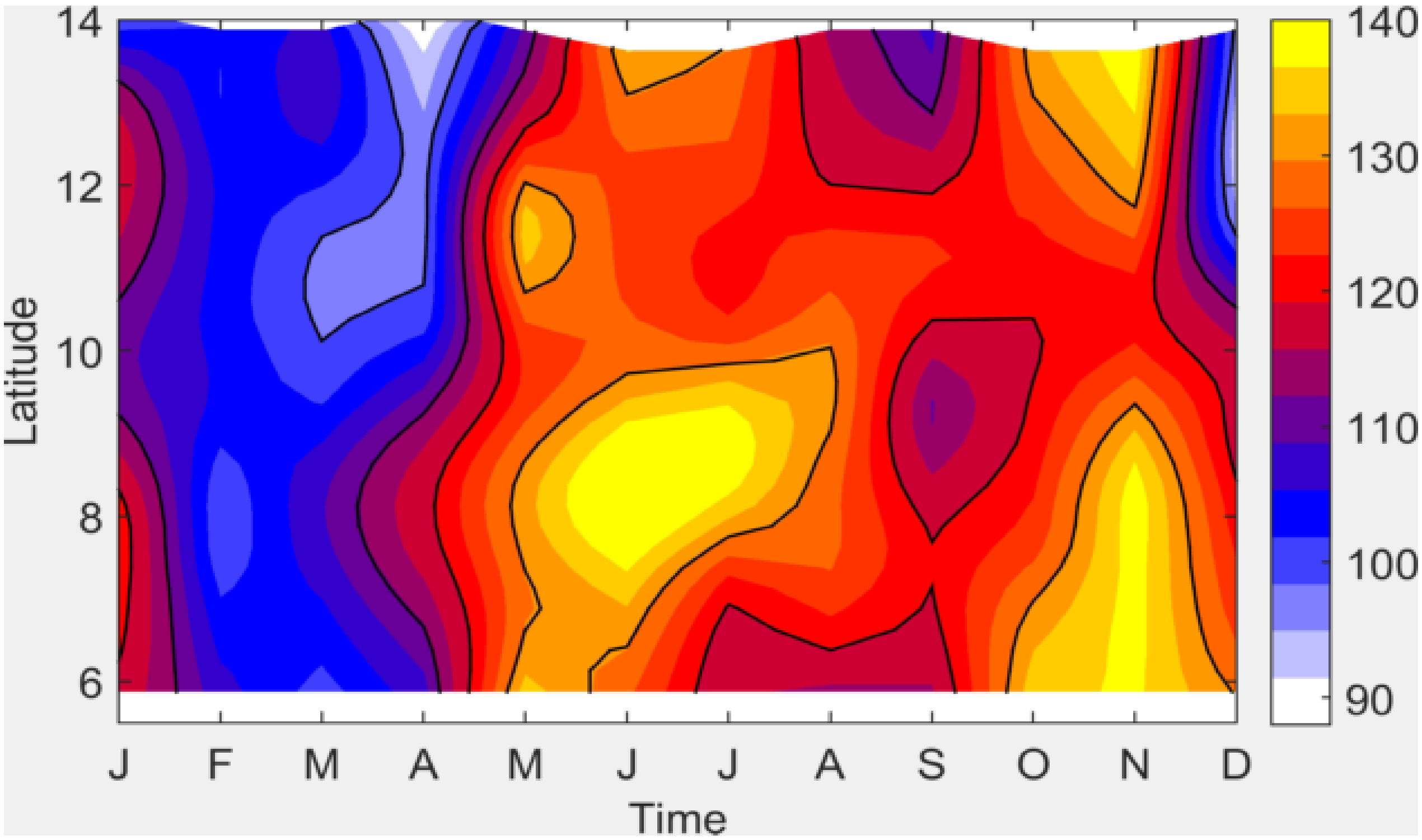

**Figure 5.** Temporal variation of depth of 20 degree isotherm (95 E to 96 E averaged) in metres.

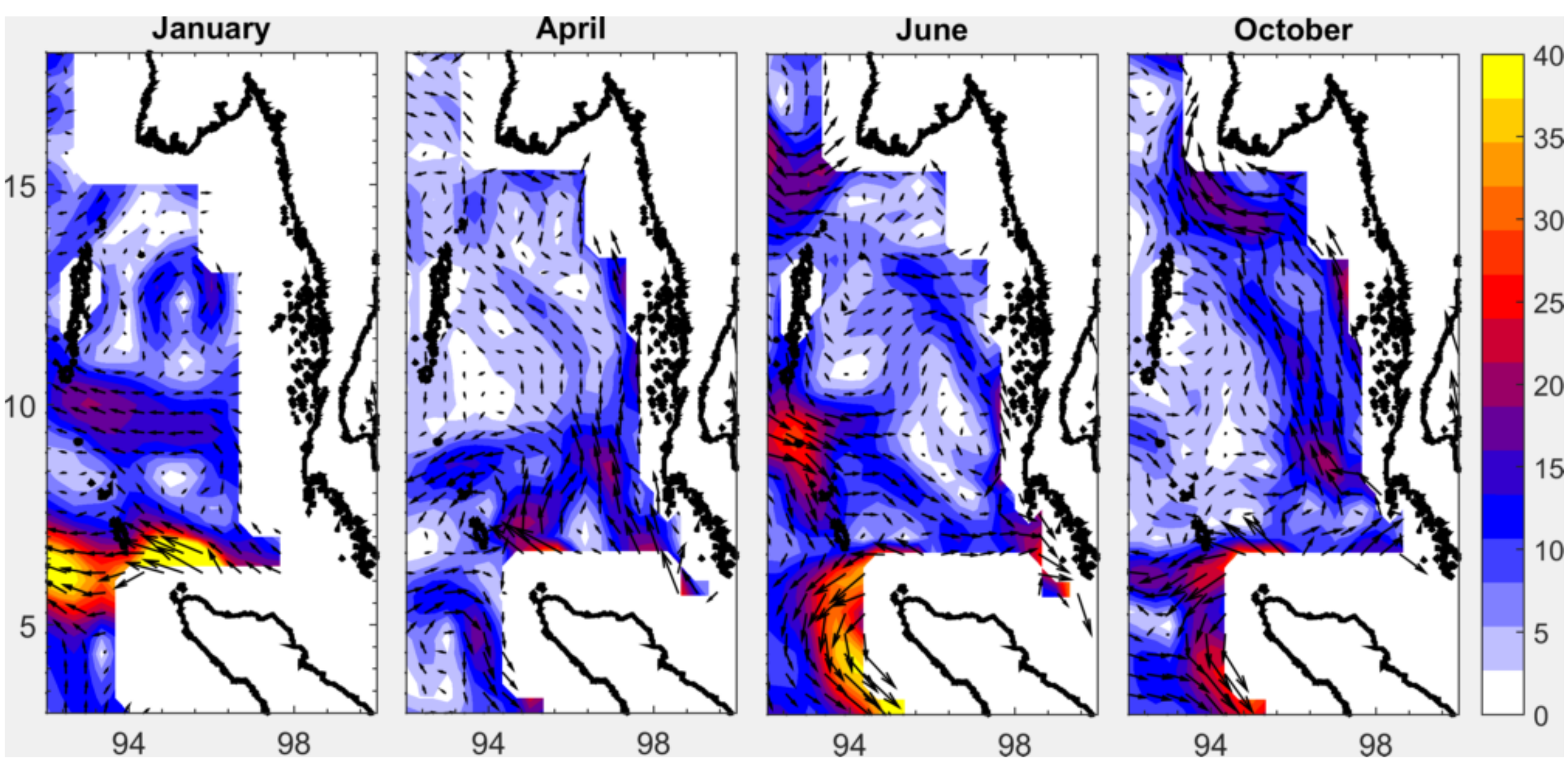

**Figure 6.** Monthly averaged OSCAR surface currents (cm/s) in January, April, June and October.

### 3.4 Transport Across Straits

AVISO-MSLA for 5 years (2005-2009) is spatially averaged to study the monthly variation of Sea Surface Height Anomalies ($SSH_{anomaly}$) of A-SEA. This is multiplied by the area of the basin to obtain the net volume of water accumulated in the region. It is observed that the water level rises in the basin between April and November (Figure 7), with maximum rate of piling up of water during April and October (marked by steep slope of the curve). The rise in SSH is attributed to the following; Rainfall, Fresh water influx from Rivers and inflow of water through the three major straits. Except the last factor, the contributions from Rainfall and Rivers are quantifiable and are hence expressed in volumes of water for comparison (Figure 7). From this, the expected influx through straits (= $SSH_{anomaly}$ – Rainfall – River Influx) could be deduced. Here, the evaporative losses are not accounted owing to its diminutive order of magnitude compared to precipitation (Previous studies[9] show that the annual mean freshwater gain (precipitation minus evaporation) of A-SEA is 120cm/year). It is found that the SSH of the basin is primarily dictated by the transport of water through the straits. The contributions from Rainfall and Rivers become substantial only during Summers. Hence, a net inward flow occurs through the straits between April and November, followed by net outward transport till March.

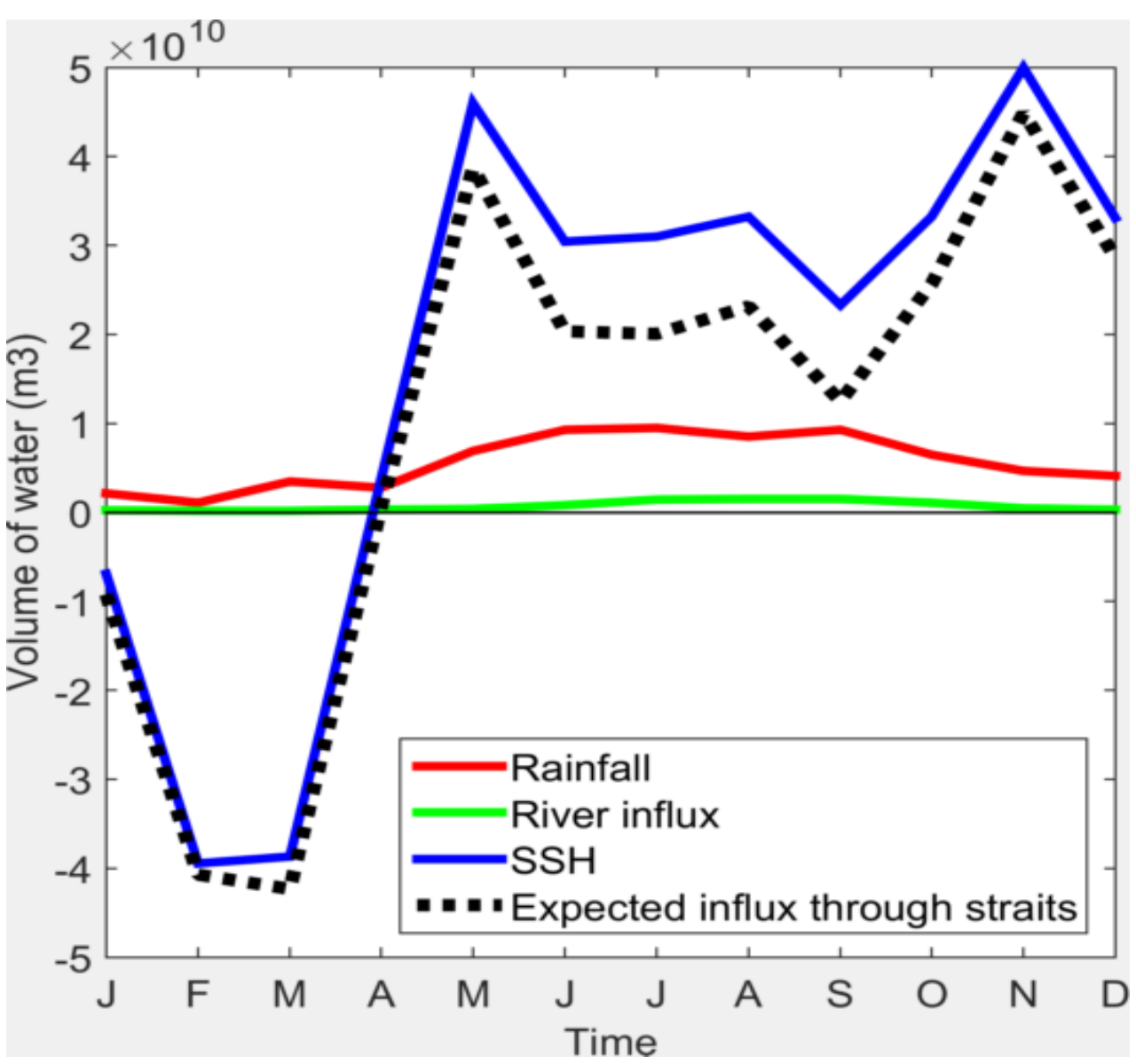

**Figure 7.** Temporal variations of the basin Rainfall, River Influx and Sea Surface Height Anomaly, expressed in volume of water.

### 3.5 Kelvin Waves

The A-SEA basin experiences very high rate of transport of water through straits in April and October (Figure 7). This is a period of equatorial Wyrtki jets[10], which hit the coast of Sumatra and reflects back as Rossby Waves and coastal Kelvin Waves. These Kelvin waves are guided along the eastern boundary of Indian Ocean and a part of this signal shall propagate into A-SEA. And the northern coast of Sumatra is the first to sense its effect. Here (within 8N), the 20 degree isotherms are found to deepen during the same period, as observed from Figure 5. This is suggestive of the downwelling nature of Kelvin waves.

The waves further propagate along the eastern boundary of A-SEA. To confirm this, monthly variations of the depth of 20 degree isotherm for longitudes 94E and 97E (averaged over latitudes 8 N and 13 N) are studied (Figure 8). The longitudes are chosen such that one represents the western part of the basin (94E) and the other along the steep continental slope on the eastern side of basin (97E). It is observed that both the longitudes experience deepening of the isotherms in April and October, but the effect is more pronounced at 97E (isotherms deepen by 30m in April and 10m in October). This is a concrete signature of downwelling in the basin and is definitely not forced locally as the winds are weaker during this period. This confirms unequivocally that the sudden burst of water into the basin through the straits, intensification of eastern boundary currents (Section 3.3) and the coincidental deepening of isotherms in April and October are the direct consequence of the propagation of downwelling Kelvin waves in A-SEA, remotely forced by equatorial Wyrtki jets.

To study the dominant modes of variability, wavelet transform of meridional mean surface currents (7N to 11N averaged) are performed, one on eastern part (95E to 97E averaged) and other along western part of the basin (92E to 94E averaged), as shown in Figures 9(a) and 9(b) respectively. The eastern region, where the effect of Kelvin waves is pronounced, elicits a strong semi-annual variability (150–230 day), which is totally absent on the western part of the domain. The occurrence of individual events of semi-annual period corresponds to October to December months of 2008, 2011 and 2012. Further, a part of spectral energy on the eastern side of the basin is confined to 40–60 day harmonics, the occurrences of which correspond to March to May months of 2008, 2011, 2012 and 2013. Hence, the Kelvin waves of A-SEA manifest as semi-annual and 40 – 60 day modes of variability.

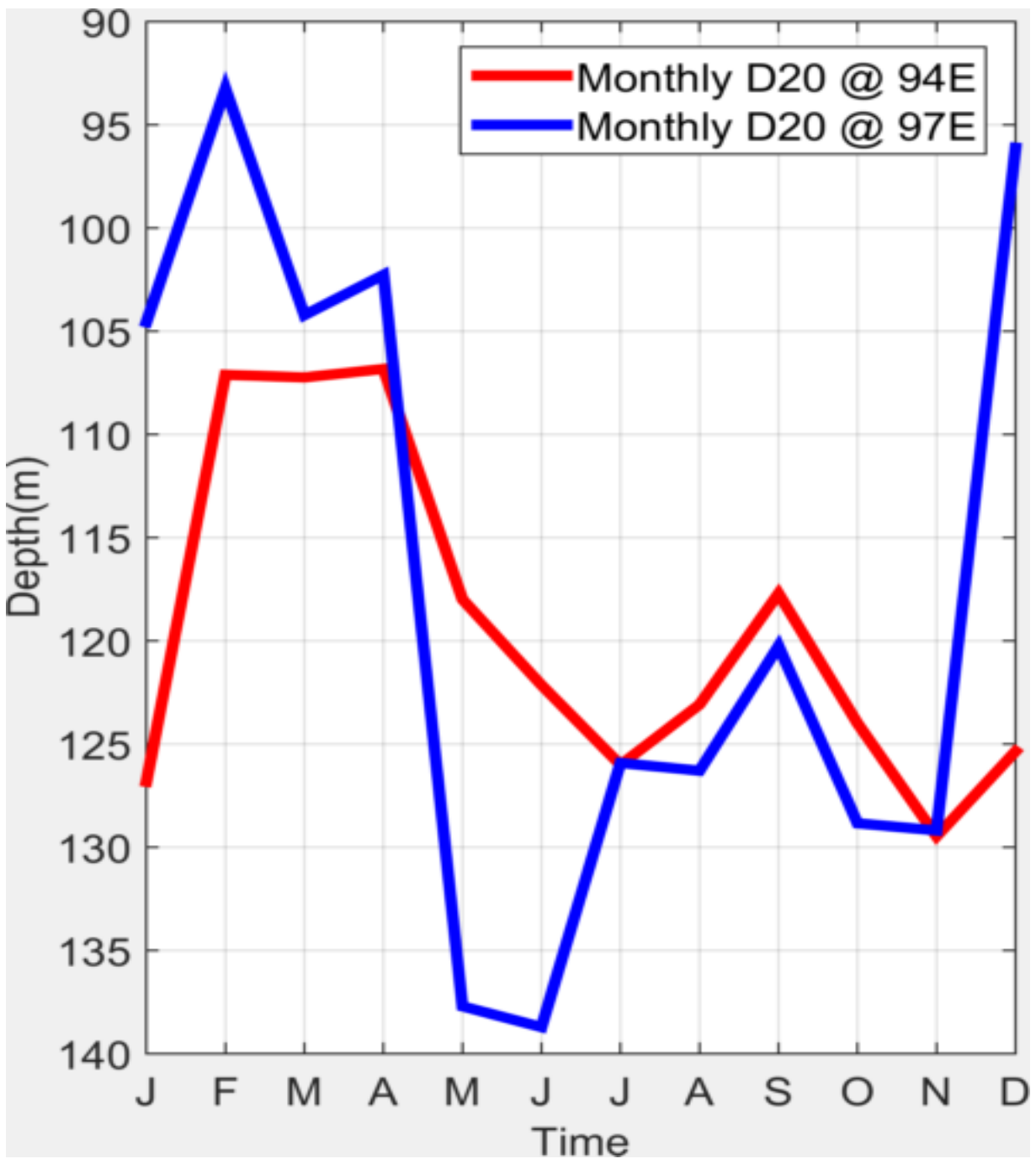

**Figure 8.** Comparison of depth of 20 degree isotherm between eastern and western regions of A-SEA.

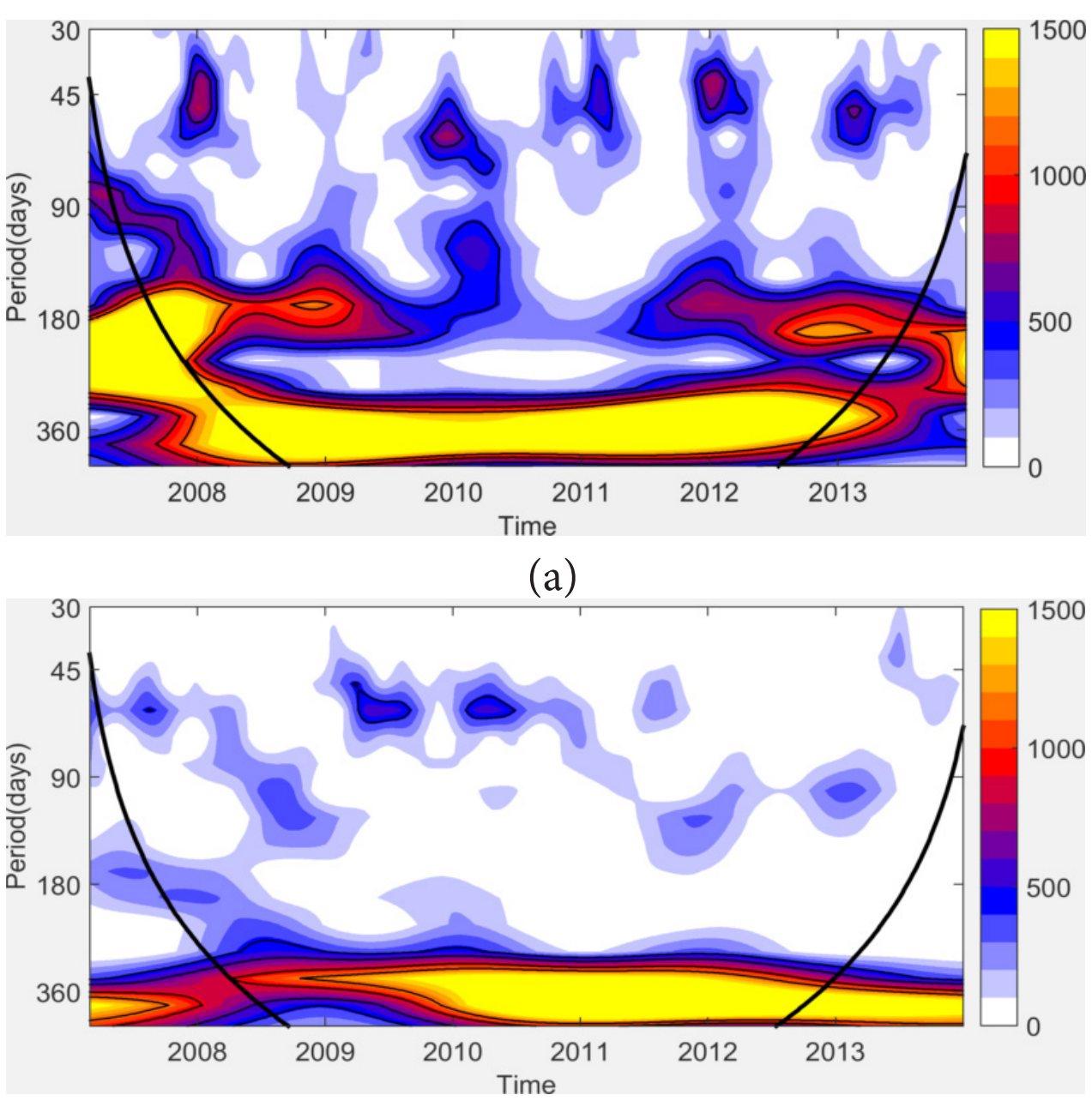

**Figure 9.** Morlet wavelet power spectra of Meridional surface currents (a) 95E to 97E zonal averaged (b) 92E to 94E zonal averaged.

## 3.6 Rossby Wave Modes

In order to identify the dominant modes of variability in MSLA, wavelet transform of AVISO-MSLA for 20 years (1995-2014) is performed. Figure 10 shows the Morlet wavelet power spectra (here, shown only from 2001 to 2006) averaged in space (93E to 97E and 7N to 12N). Annual (360 days), semi-annual (180 days) and 120-day harmonics are the most dominant modes of variability present in MSLA. The annual mode corresponds to a cycle of rise (during Summer) and fall (during Winter) of SSH per year, controlled by transport across straits (Section 3.4).

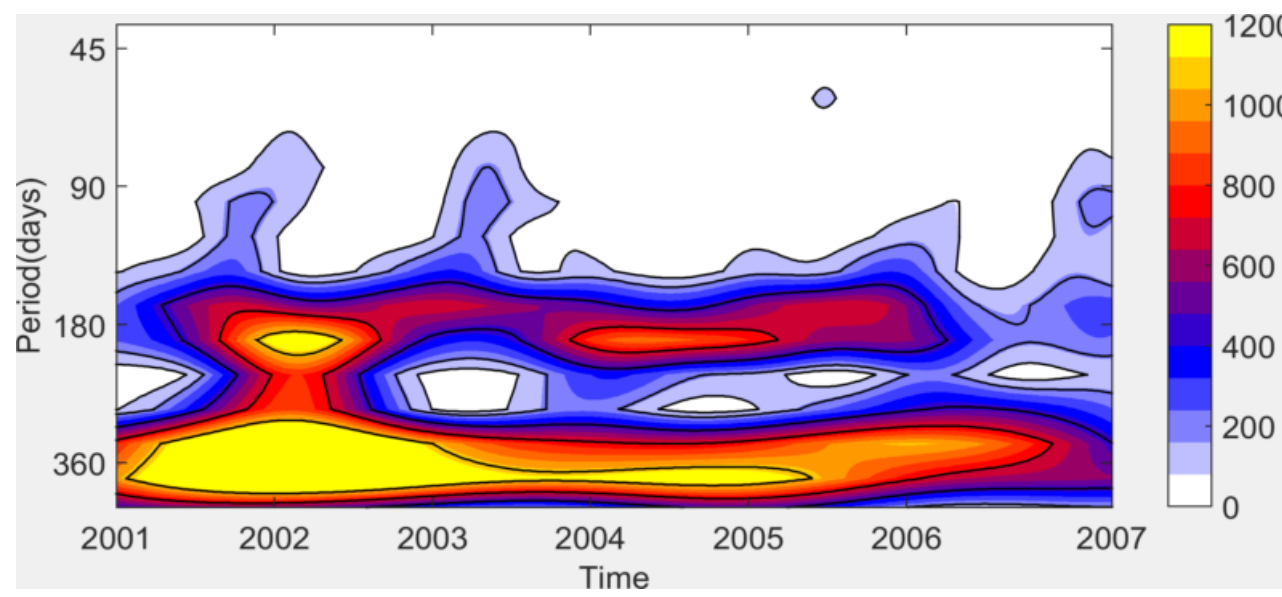

**Figure 10.** Morlet wavelet power spectra for AVISO-MSLA.

### *3.6.1 180-Day Variability*

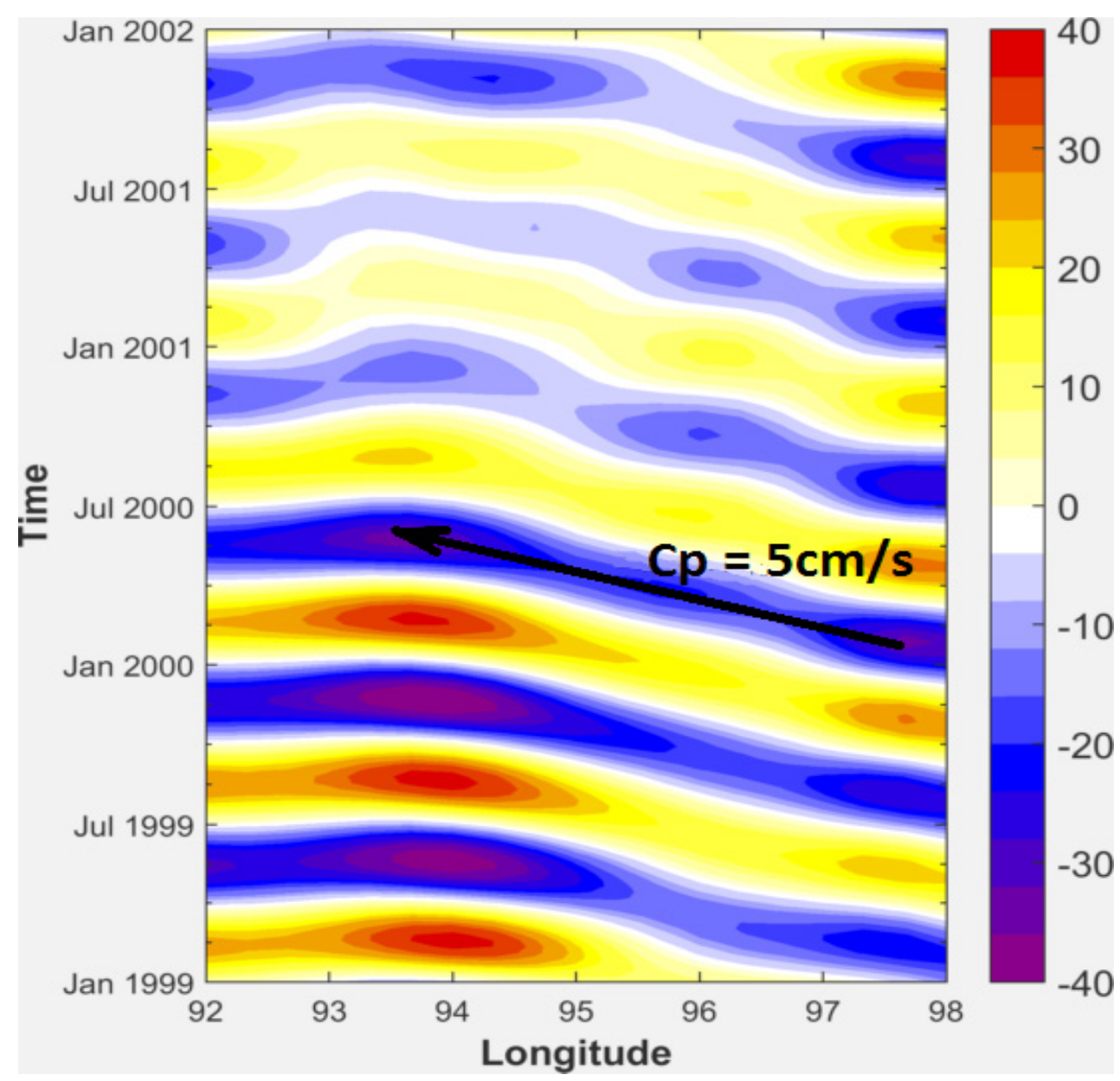

**Figure 11.** Longitude-Time plot of 150-230 day bandpass-filtered MSLA.

The spectral energy spread over a band of 150-230 days period is filtered for analysis. Figure 11 shows the Hovemuller diagram (Longitude-Time plot) of the filtered MSLA for the latitudinal band 7N to 12N. Clear signatures of phase propagation of Rossby Waves with speed Cp = 5cm/s and $(L_x K_x/2\pi) = 0.5$ is observed, where $L_x$ and $K_x$ are the length of the basin and wavenumber respectively in the zonal direction. Figure 12 shows the evolution of MSLA of the basin, overlaid with OSCAR currents (both 150-230 days Band-passed) from July 2011 to June 2012 (15th of every month). It shows that the domain is composed of vortices; dome-like (and dish-like) structures of high (and low) SSH, with anti-cyclonic (and cyclonic) circulation around them. These vortices propagate north-westward throughout the year, the northward component of which is rendered by the mean flow. These are characteristic phase propagations of Rossby waves, which manifest as cyclonic and anti-cyclonic "persistent and periodic" gyres in the general circulation (Section 3.3). Hence, these vortices are not the consequence of intense momentum flux through straits, rather the result of internal perturbations (locally forced by winds and by remotely forced Kelvin waves) propagating as Rossby waves.

### 3.6.2 120-Day Variability

Figure 13 shows the Hovemuller diagram of the 100-140 day bandpass-filtered MSLA for the latitudinal band 7N to 12N. Definite signatures of Rossby wave reflection is observed here. A wave packet of westward Group velocity (Cg = 1.3 cm/s) and westward Phase speed (Cp = 13 cm/s) strikes the coasts of ANI and reflect as another wave packet of eastward Group velocity (Cg = 0.6 cm/s) and westward Phase speed (Cp = 6 cm/s). The incident wave is observed to have longer wavelength $[(L_x K_x/2\pi) = 0.5]$ than the reflected wave $[(L_x K_x/2\pi) = 1.5]$, which is the characteristic of Rossby waves.

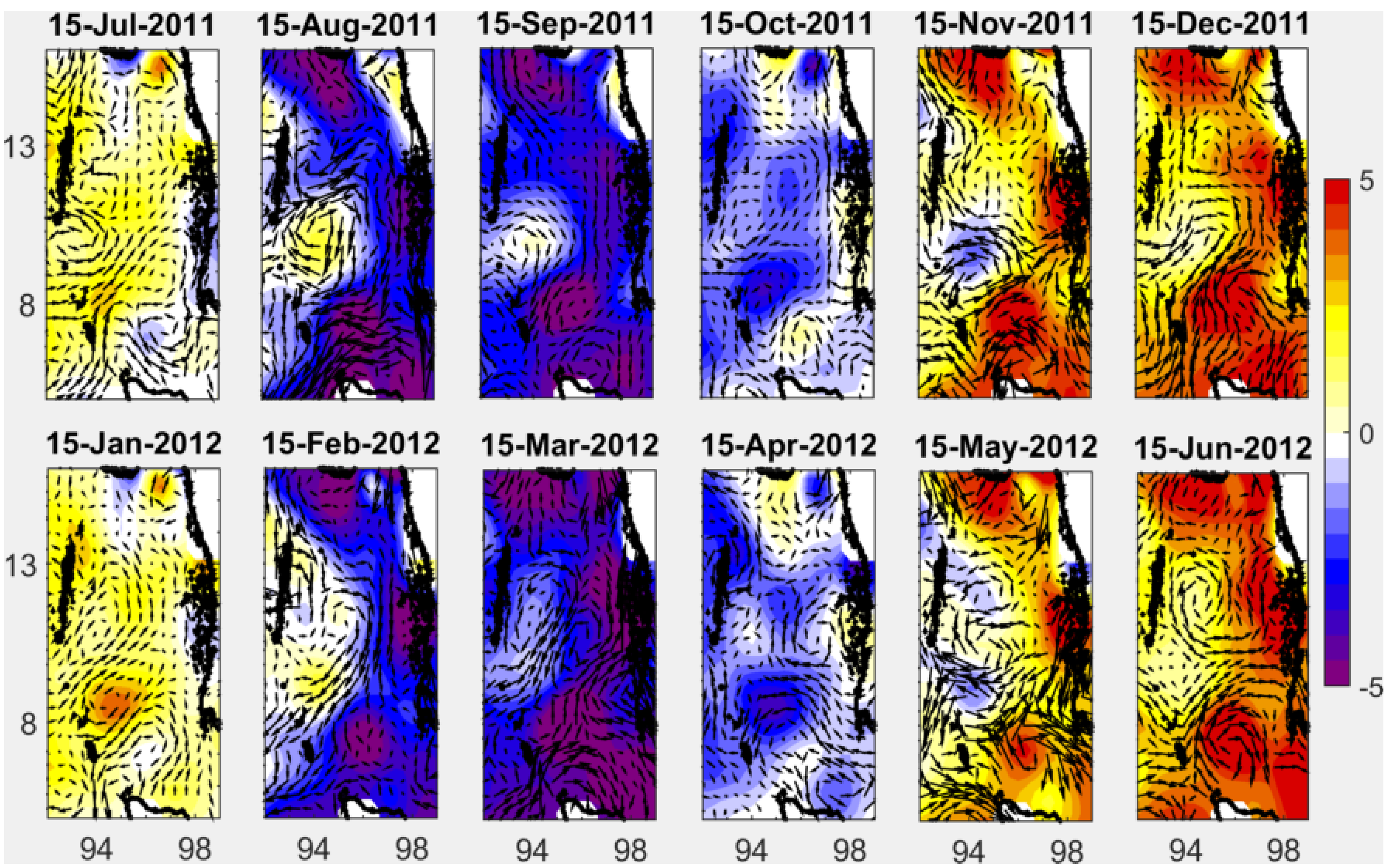

**Figure 12.** AVISO-MSLA overlaid with OSCAR currents (150-230 day bandpass-filtered).

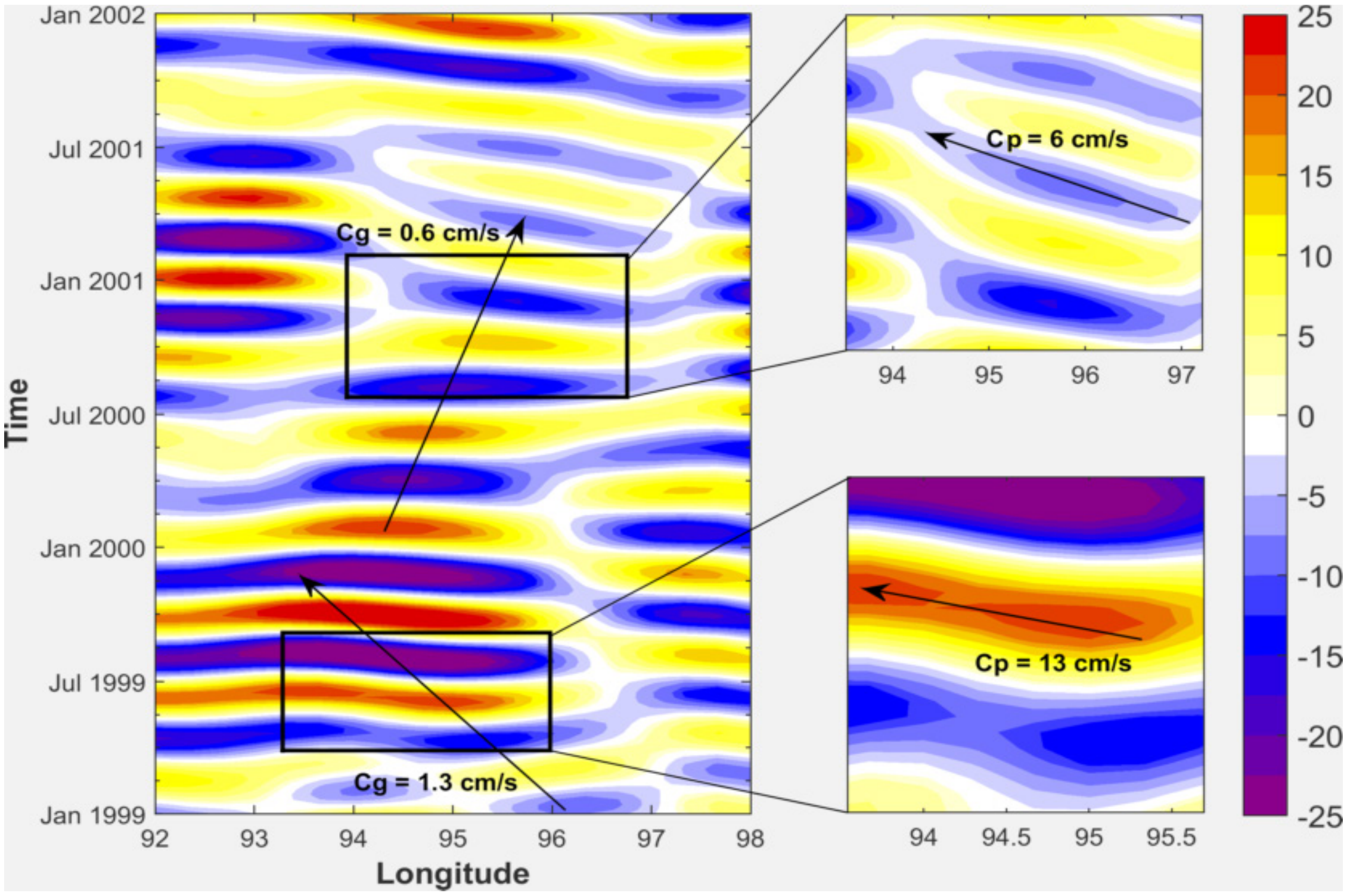

**Figure 13.** Longitude-Time plot of 100-140 day bandpass-filtered MSLA.

## 4. Conclusions

This study delves into the dynamics of Andaman Sea circulation with specific insight into dominant wave modes of the basin. The following are the major conclusions arrived at by this study.

- Wind system over Andaman Sea conforms to the general seasonal variability of Indian Ocean. They are strong south-westerlies during Summers (May to September) and relatively weaker north-easterlies during Winters (November to March).
- A cell of strong downwelling forms to the north of Indonesia during Summer. This is accompanied by a high value of negative Ekman pumping velocity over the same region. This means that the effect of local forcing is predominant in Summers.
- Sea Surface Height of Andaman Sea is primarily influenced by the transport across the straits between the islands of Andaman and Nicobar. The flow through the straits is inward from April to November, followed which the flow becomes truly outward from December to March.
- During the months of April and October, coastal Kelvin waves propagate along the eastern boundary of Andaman Sea, remotely forced by equatorial Wyrtki jets. This is corroborated by intensification of alongshore meridional surface currents, abrupt burst of water into the basin through the straits and the concomitant deepening of 20 degree isotherms. The basin experiences downwelling along its trajectory. The semi-annual and 40-60 day harmonics are the dominant modes of variability associated with the wave.
- Dominant modes of Rossby waves in Andaman Sea are 180-day and 120-day harmonics. Rossby waves of semi-annual period propagate north-westward with phase speed of 5cm/s and are responsible for the vortex formation in the domain. These appear as cyclonic and anticylonic gyres or circulation with alternating polarity at the north and south of the basin. The 120-day mode manifest as long Rossby waves with westward Group velocity, which strikes the coasts of Andaman and Nicobars and reflects back as short Rossby waves of eastward Group velocity.

## 5. Acknowledgement

I thank the faculty of Centre for Atmospheric and Oceanic Sciences, IISc for fruitful discussions and valuable input for the initiation of this work. I thank my parents and colleagues for the support and encouragement.